\documentclass[aps,prl,twocolumn,superscriptaddress]{revtex4-1}

\usepackage{siunitx}
\DeclareSIUnit{\au}{{a.u.}}
\DeclareSIUnit{\cm}{{cm}}
\DeclareSIUnit{\mbar}{{mbar}}

\usepackage{textcomp}
\usepackage{float}
\usepackage{graphicx}% Include figure files

\usepackage{fp}
\newcount\boxheight
\newcount\boxwidth

\newcommand\testaspectone[1]{%
	\setbox0=\hbox{#1}%
	\boxheight=\ht0\relax%
	\boxwidth=\wd0\relax%
	\FPdiv\theaspect{\the\boxheight}{\the\boxwidth}%
	\FPmul\theaspect{\theaspect}{150}%
	\FPadd\theaspect{\theaspect}{20}%
	\FPeval\theaspect{round(\theaspect:1)}
	\copy0%
}

\newcommand\testaspecttwo[1]{%
	\setbox0=\hbox{#1}%
	\boxheight=\ht0\relax%
	\boxwidth=\wd0\relax%
	\FPdiv\theaspect{\the\boxheight}{\the\boxwidth}%
	\FPmul\theaspect{\theaspect}{600}%
	\FPadd\theaspect{\theaspect}{40}%
	\FPeval\theaspect{round(\theaspect:1)}
	\copy0%
}

\usepackage{booktabs}

\usepackage{xspace} % To get the spacing after macros right
\usepackage{amssymb}

\usepackage{mathtools, nccmath}

\newcommand{\ie}{i.\,e.}

\newcommand{\ExB}{$ \hat{E} \times \hat{B}$\xspace} 

\usepackage[hidelinks]{hyperref}

\begin{document}
	
	%TC:macrocount \si [inlinemath]
	%TC:macrocount \ref [word]
	%TC:macrocount \ExB [word]
	%TC:macrocount \eg [word]
	%TC:macrocount \ie [word]
	%TC:macrocount \comsol [word]
	%TC:macrocount \geant [word]

%\preprint{}

%TC:ignore 
%\title{Muon cooling: Transverse Compression}
\title{Demonstration of Muon-Beam Transverse Phase-Space Compression}


\author{A. Antognini}
\email[]{aldo@phys.ethz.ch}
\affiliation{Institute for Particle Physics and Astrophysics, ETH Z\"urich, 8093 Z\"urich, Switzerland}
\affiliation{Paul Scherrer Institute, 5232 Villigen-PSI, Switzerland}

\author{N. J. Ayres}
\affiliation{Institute for Particle Physics and Astrophysics, ETH Z\"urich, 8093 Z\"urich, Switzerland}

\author{I. Belosevic}
\email[]{ivanabe@phys.ethz.ch}
\affiliation{Institute for Particle Physics and Astrophysics, ETH Z\"urich, 8093 Z\"urich, Switzerland}
%\thanks{}

\author{V. Bondar}
\affiliation{Institute for Particle Physics and Astrophysics, ETH Z\"urich, 8093 Z\"urich, Switzerland}

\author{A. Eggenberger}
\affiliation{Institute for Particle Physics and Astrophysics, ETH Z\"urich, 8093 Z\"urich, Switzerland}

\author{M. Hildebrandt}
\affiliation{Paul Scherrer Institute, 5232 Villigen-PSI, Switzerland}

\author{R. Iwai}
\affiliation{Institute for Particle Physics and Astrophysics, ETH Z\"urich, 8093 Z\"urich, Switzerland}

 \author{D. M. Kaplan}
 \affiliation{Illinois Institute of Technology, Chicago, IL 60616 USA}

\author{K. S. Khaw}
\altaffiliation{\textit{Present address}: Tsung-Dao Lee Institute, and School of Physics and Astronomy, Shanghai Jiao Tong University, Shanghai 200240, China}
\affiliation{Institute for Particle Physics and Astrophysics, ETH Z\"urich, 8093 Z\"urich, Switzerland}

\author{K. Kirch}
\affiliation{Institute for Particle Physics and Astrophysics, ETH Z\"urich, 8093 Z\"urich, Switzerland}
\affiliation{Paul Scherrer Institute, 5232 Villigen-PSI, Switzerland}

\author{A. Knecht}
\affiliation{Paul Scherrer Institute, 5232 Villigen-PSI, Switzerland}

\author{A. Papa}
\affiliation{Paul Scherrer Institute, 5232 Villigen-PSI, Switzerland}
\affiliation{Dipartimento di Fisica, Universit\`a di Pisa, and INFN sez. Pisa, Largo B. Pontecorvo 3, 56127 Pisa, Italy}

\author{C. Petitjean}
\affiliation{Paul Scherrer Institute, 5232 Villigen-PSI, Switzerland}

\author{T. J. Phillips}
\affiliation{Illinois Institute of Technology, Chicago, IL 60616 USA}

\author{F. M. Piegsa}
\altaffiliation{\textit{Present address}: Laboratory for High Energy Physics, Albert Einstein Center for Fundamental Physics, University of Bern, CH-3012 Bern, Switzerland}
\affiliation{Institute for Particle Physics and Astrophysics, ETH Z\"urich, 8093 Z\"urich, Switzerland}

\author{N. Ritjoho}
\affiliation{Paul Scherrer Institute, 5232 Villigen-PSI, Switzerland}

\author{A. Stoykov}
\affiliation{Paul Scherrer Institute, 5232 Villigen-PSI, Switzerland}

\author{D. Taqqu}
\affiliation{Institute for Particle Physics and Astrophysics, ETH Z\"urich, 8093 Z\"urich, Switzerland}

\author{G. Wichmann}
\altaffiliation{\textit{Present address}: Laboratory of Physical Chemistry, ETH Z\"urich, 8093 Z\"urich, Switzerland}
\affiliation{Institute for Particle Physics and Astrophysics, ETH Z\"urich, 8093 Z\"urich, Switzerland}

\collaboration{muCool collaboration}

%\date{\today}

\begin{abstract}
We demonstrate efficient transverse compression of a 12.5~MeV/c muon beam stopped in a helium gas target featuring a vertical density gradient and crossed electric and magnetic fields.
% The 14\,mm wide muon stopping distribution was reduced to 0.25\,mm (RMS) within 3.5~\si{\micro \s}.
The muon stop distribution extending vertically over 14\,mm was reduced to a 0.25\,mm size (RMS) within 3.5~\si{\micro \s}.
% 
% The vertical size of the muon stopping distribution was reduced from 14 to 0.25~mm (RMS) within 3.5~\si{\micro \s}. 
The simulation including cross sections for low-energy $\mu^+$-$\text{He}$ elastic and  charge exchange ($\mu^+\leftrightarrow $ muonium) collisions describes the measurements well. By combining the transverse compression stage with a previously demonstrated longitudinal compression stage, we can improve the phase space density of a $\mu^+ $ beam by a factor of $ 10^{10} $ with $ 10^{-3} $ efficiency.

\end{abstract}

%\keywords{}

\maketitle
%TC:endignore 

Next generation precision experiments with muons and muonium atoms~\cite{Gorringe2015}, such as muon $ g-2 $ and EDM measurements~\cite{Iinuma2011, Adelmann2010, Crivellin2018}, muonium spectroscopy~\cite{Cr}, and muonium gravity measurements~\cite{Kirch2014a,Kaplan2018}, require high-intensity muon beams at low energy with small transverse size and energy spread. No muon beams currently available fulfill these requirements. To improve the quality of the muon beam, phase space cooling techniques are needed. Conventional methods, such as stochastic cooling~\cite{VanderMeer1985} and electron cooling~\cite{Budker1978}, are not applicable due to the short muon lifetime of $ 2.2 $~\si{\micro \s} (at rest). Alternative beam cooling techniques based on muon energy moderation in materials have been developed~\cite{Muhlbauer1999,Morenzoni1994}, however, they suffer from low cooling efficiencies ($<10^{-4} $).

\begin{figure*}[t!]
%	\testaspectone{
	\testaspecttwo{
		\includegraphics[width=1.0\textwidth]{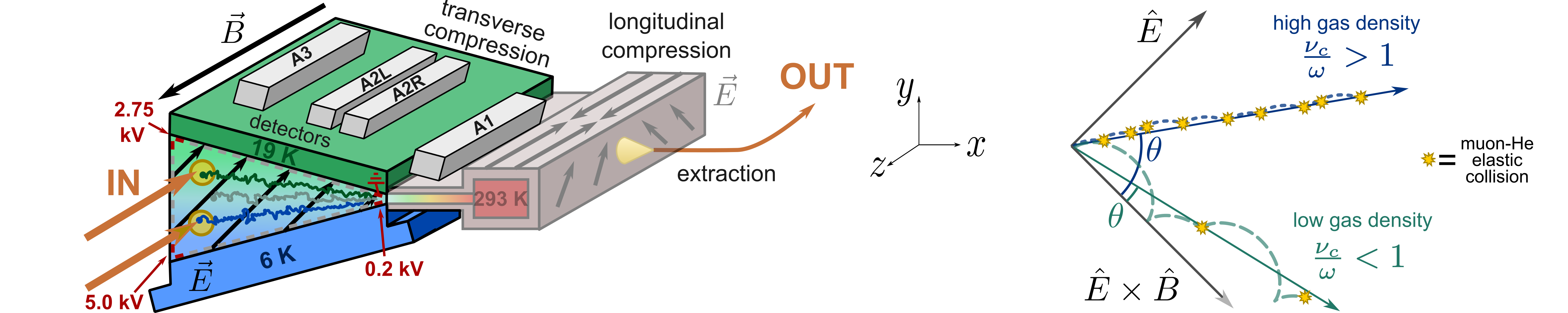}
	}

% \theaspect~words
	\caption{\label{scheme}
		(Left) Schematic diagram of the muCool device. A standard muon beam is stopped in a cryogenic helium gas target with a vertical temperature gradient inside a 5~T field. The extent of the stopped muons is reduced first in the transverse ($y$), then in the longitudinal ($z$) direction. The now sub-mm muon beam is extracted through an orifice into vacuum and re-accelerated along the $ z $-axis. The electrodes used to define the required electric field in the transverse compression stage are also sketched (gray and red rectangles), along with the values of applied high voltage. (Right) Schematic of $ \mu^+ $ trajectories in various gas densities and in crossed electric and magnetic fields.
	} 
\end{figure*}

At the Paul Scherrer Institute, we are developing a novel device that reduces the full (transverse and longitudinal) phase space of a surface~\cite{Pifer1976a} (28~MeV/c momentum) $\mu^{+}$ beam by 10 orders of magnitude with $10^{-3}$ efficiency~\cite{Taqqu2006}. 
This so-called muCool device is placed inside a $ 5 $~\si{\tesla} magnetic field, pointing in the $ +z $-direction (Fig.~\ref{scheme}).
First, a surface muon beam propagating in the $-z $-direction is stopped in a few mbar of cryogenic helium gas (at 10~K on average), reducing the muon energy from MeV to the eV range. 
% This initial slowing down reduces muon beam momentum spread in $ x $, $ y $ and $ z $-direction by a factor of 100.
As a result of the slowing down, the momentum spreads in all directions $\sigma_{p_x,p_y,p_z}$ are reduced from $\mathcal{O}$(MeV/c) to about $20 $~\si{\keV/c}.
The 6D phase space is subsequently further reduced by compressing the spatial extent $\sigma_{x,y,z}$ (RMS) while keeping the same momentum spread.
% To compress the full 6D phase space, the reduction of the momentum spread needs to be accompanied by the reduction of the beam spot size. 
This is achieved by guiding the stopped muons into a sub-mm spot using a combination of strong electric and magnetic fields and gas density gradients in three stages (see Fig.~\ref{scheme}).

In the first stage (transverse compression), the electric field is perpendicular to the magnetic field and at $ 45 $\si{\degree} with respect to the $ x $-axis: $ \vec{E} = (E_x, E_y, 0) $, with $ E_x=E_y\approx 1$~\si{\kV/\cm}. In vacuum, such fields would prompt the stopped muons to drift in the \ExB-direction, performing cycloidal motion at the cyclotron frequency $\omega=eB/m_{\mu}$, where $ m_{\mu} $ is the muon mass.

However, in the muCool device, the muons collide with He gas atoms with an average frequency $ \nu_{c} $, which depends on the gas density, elastic $ \mu^+ $-He cross section, and muon energy.
These collisions lead to muon energy loss and direction change, modifying the muon motion compared to that in vacuum, so that muons drift at an angle $ \theta $ relative to the \ExB-direction~\cite{Heylen1980}:
% The collisional energy loss is balanced by the acceleration due to the electric field, thus maintaining the muon energy and energy spread in the eV range. 
% The muon direction change in the collisions 
% 
%
\begin{equation}\label{eq:drift_angle}
\tan\theta= \frac{\nu_c}{\omega}.
\end{equation}
This equation can be understood as follows: for $ \nu_{c}>\omega $, muons collide with He atoms at the beginning of the cycloidal motion, before their trajectory is significantly bent by the magnetic field, resulting in a large $ \theta $ (blue trajectory in Fig.~\ref{scheme}). For $ \nu_{c} < \omega $, muons travel further in the \ExB-direction between collisions (similar to vacuum), resulting in a smaller $ \theta $ (green trajectory in Fig.~\ref{scheme}).

%muon doesn't travel far in the \ExB-direction on/along the cycloidal trajectory/path.

%The reason for such dependence on $ \nu_{c} $ is as follows: for $ \nu_{c}>\omega $, the muon does not complete the full period of the cycloidal motion before the next collision (blue trajectory in Fig.~\ref{scheme}), resulting in a large $ \theta $. If $ \nu_{c}<\omega $, the muon performs several periods of cycloidal motion between collisions, resulting in a smaller $ \theta $ (green trajectory in Fig.~\ref{scheme}).

By changing the collision frequency we can thus manipulate the muon drift direction.
In the transverse compression stage, we change the collision frequency by modifying the gas density: the top of the apparatus is kept at higher temperature than the bottom (19~K versus  6~K), creating a temperature gradient in the vertical ($ y $-) direction. The pressure is chosen so that $ \frac{\nu_c}{\omega}=1 $ at $ y=0 $. Muons at different $ y $-positions (at fixed $ x $-position) experience different densities, resulting in different drift directions: a muon stopped at the bottom of the target (higher density) experiences more collisions ($ \frac{\nu_c}{\omega}>1 $) and will thus drift predominately in the $ \hat{E} $-direction (large $ \theta $, upwards), while a muon stopped in the top part (lower density) collides less frequently with He atoms ($ \frac{\nu_c}{\omega}<1 $), resulting in drift predominately in the \ExB-direction (small $ \theta $, downwards). As a result, muons stopped at different $ y $-positions converge towards $ y=0 $, while drifting in the $ +x $-direction, ultimately reaching a size $\sigma_y\approx 0.25$~mm. Throughout this drift, the collisional energy loss is balanced by the acceleration in the electric field, so that $ \sim 20 $~\si{\keV/c} momentum spreads $\sigma_{p_x,p_y,p_z}$ are maintained; combined with the spatial compression in the $y$-direction we thus obtain transverse ($\sigma_y \cdot \sigma_{p_y}$) phase-space compression.

Subsequently, the muons enter the second stage, at room temperature, where $\vec{E}=(0,E_y, \pm E_z)$, with $E_y=2E_z = 0.1$~kV/cm.
The $E_z$-component drifts the muons towards $z=0$, thus reducing their $\sigma_z$ extent to sub-mm (longitudinal compression). Simultaneously, at this low density, the $E_y $-component drifts the muons in the \ExB ($ +x $) direction, towards the extraction stage. From there, the sub-mm muon beam can be extracted though a small orifice into vacuum, re-accelerated along the $ z $-direction with pulsed electric fields to keV energies, and extracted from the magnetic field. 
% The final muon beam extent in the $ x $-direction is determined by the muon \ExB drift velocity and the pulsing frequency: for 1~kV/cm drift field and 20 MHz acceleration frequency it is 1\,mm.
Efficient muon-beam longitudinal compression, including an \ExB-drift, has already been demonstrated~\cite{Bao2014,Belosevic2019}.

In this Letter we demonstrate the muon transverse compression stage of the muCool device.
For this demonstration, about $2\cdot10^4~\mu^+$/s at 12.5~MeV/c were injected into the $ 25 $-cm-long target placed inside a 5-tesla solenoid. 
Before entering the target, the muons traversed a 55-\si{\micro \meter}-thick entrance detector, several thin foils, and a copper aperture, defining the beam injection position. 
 
The transverse target gas volume was enclosed by a Kapton foil, folded and glued around two triangular PVC end-caps. 
% Kapton and PVC are both electrical and thermal insulators, capable of sustaining high voltage, while keeping the heat transport between top and bottom walls small. 
The large thermal conductivity of single crystal sapphire plates glued to the top and bottom target walls assured homogeneous wall temperatures. The required He gas temperature gradient from 6 to 19~K was produced by heating the top sapphire and thermally connecting the bottom sapphire to a cold finger~\cite{Wichmann2016a}. 

The Kapton foil enclosing the gas volume was lined with electrodes extending along the $ z $-direction. The 45\si{\degree} electric field was defined by applying appropriate voltages to several of these electrodes (see Fig.~\ref{scheme}), which were connected to other electrodes via voltage dividers.

Several plastic scintillator bars (A1...A3) were placed around the target to monitor the muon movement by detecting the $\mu^+$-decay positrons. The scintillators' position resolution was improved by embedding them in collimators, mechanically decoupled from the target.
%Collimation around the scintillators improved their position resolution.
An additional downstream scintillator system (back detector) was used for aligning the magnetic field and the beam on the target axis (the muon beam follows the magnetic field lines). 
Wavelength-shifting fibers, glued inside scintillator grooves, transported the scintillation light from cryogenic to room temperature, where it was read out using silicon photomultipliers.
% The detectors consisted of with a groove, inside which a wavelength-shifting fiber was glued. These 2~m-long fibers transported the scintillation light from the cryogenic temperatures in vacuum to room temperature and air, where they were read out by $1.3 \times 1.3$~\si{\mm \squared} silicon photomultipliers. 

\begin{figure}[b!]
	\testaspectone{
		
				\includegraphics[width=1.0\columnwidth]{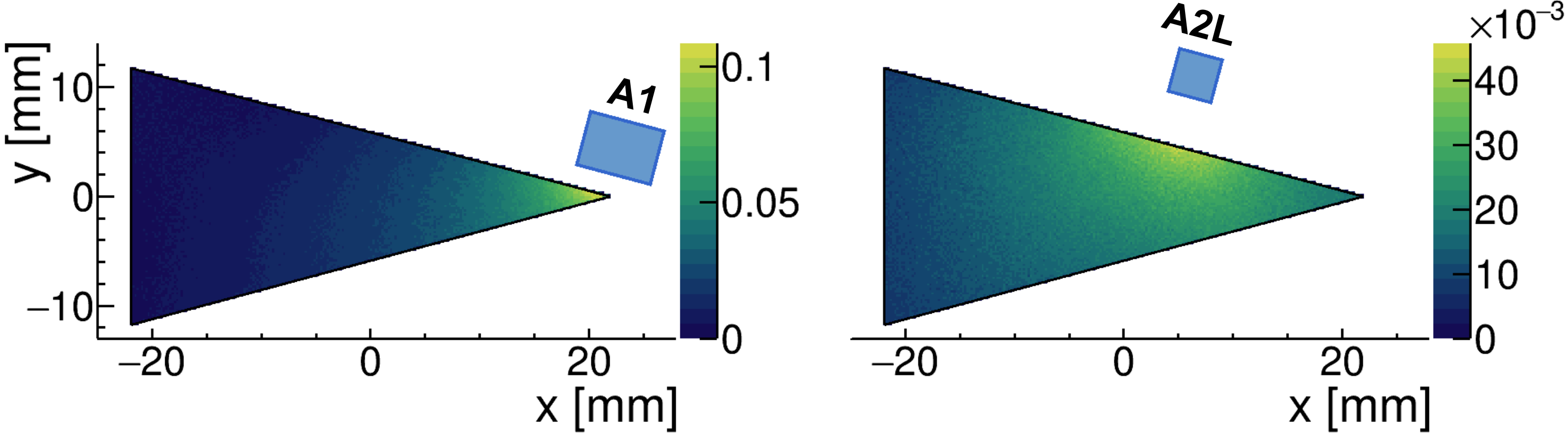}

}

% \theaspect~words
	\caption{Projection of the detection efficiency of the A1 and A2L positron detectors in the $ xy $-plane, simulated with Geant4~\cite{Agostinelli2003}. Colorscale represents positron detection efficiency. \label{fig:trans_detector_acc}}
\end{figure}
The probability of detecting the decay positrons versus the muon decay position is shown in Fig.~\ref{fig:trans_detector_acc} for detectors A1 and A2L.
By recording time spectra for each detector (with $ t=0 $ given by the muon entrance detector), we can indirectly observe the muon drift. To demonstrate transverse compression we need to show that the muon beam vertical size decreases during the drift.
This is achieved by comparing the time spectra obtained when the temperature gradient ($ 6 $--$ 19 $~K at $ 8.6 $~mbar) is applied, to the time spectra with negligible temperature gradient ($ 4 $--$ 6 $~K at $ 3.5 $~mbar): in the first case both drift and compression are expected; in the second basically only drift (``pure drift''). 
We further increased the contrast between ``compression'' and ``pure-drift'' measurements by injecting the muons through either ``top-hole'' or ``bottom-hole'' apertures at $y=\pm4.5$~mm (see Fig.~\ref{scheme}). If we were to inject the muons through a single large aperture, most muons would stop close to $y=0$, and would thus drift straight towards the target tip for both ``compression'' and ``pure drift.'' Contrarily, with either of the two smaller apertures, we target a narrow gas region, with distinct density conditions for ``compression'' and ``pure drift.''

\begin{figure}[t!]
	\testaspectone{
		
		\includegraphics[width=1.0\columnwidth]{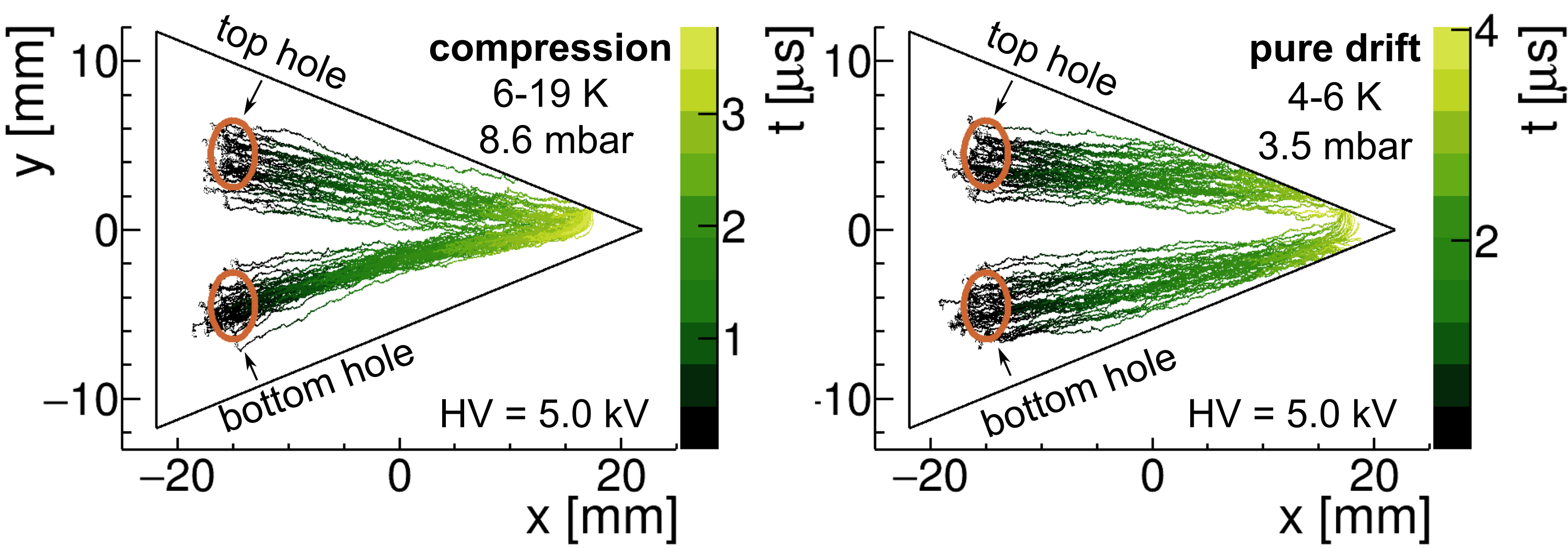}

	} 					

% \theaspect~words
	\caption{Muon trajectories in $ xy $-plane simulated with Geant4 for ``top-hole'' and ``bottom-hole'' injection and for ``compression'' (left) and ``pure-drift'' (right) density conditions. Small amount of compression visible in ``pure drift'' simulation is caused by residual temperature gradient. \label{fig:trajectories}}
	
\end{figure}

Geant4~\cite{Agostinelli2003} simulations of the muon trajectories for various experimental configurations are shown in Fig.~\ref{fig:trajectories}. The simulation included the most relevant low-energy $\mu^+$-$\text{He}$ interactions: elastic collisions and charge exchange. The needed cross sections were appropriately scaled from the proton-He cross sections~\cite{Belosevic2020, Taqqu2006}. Simulations show that for the ``compression'' case, muons injected through both apertures reach the tip of the target efficiently, while in the ``pure drift'' case, most of the muons stop in the target walls before reaching the tip.

\begin{figure}[t!]
	%			\caption*{\textbf{``top hole'', with misalignment}}
	\testaspectone{	

		\includegraphics[width=1.0\columnwidth]{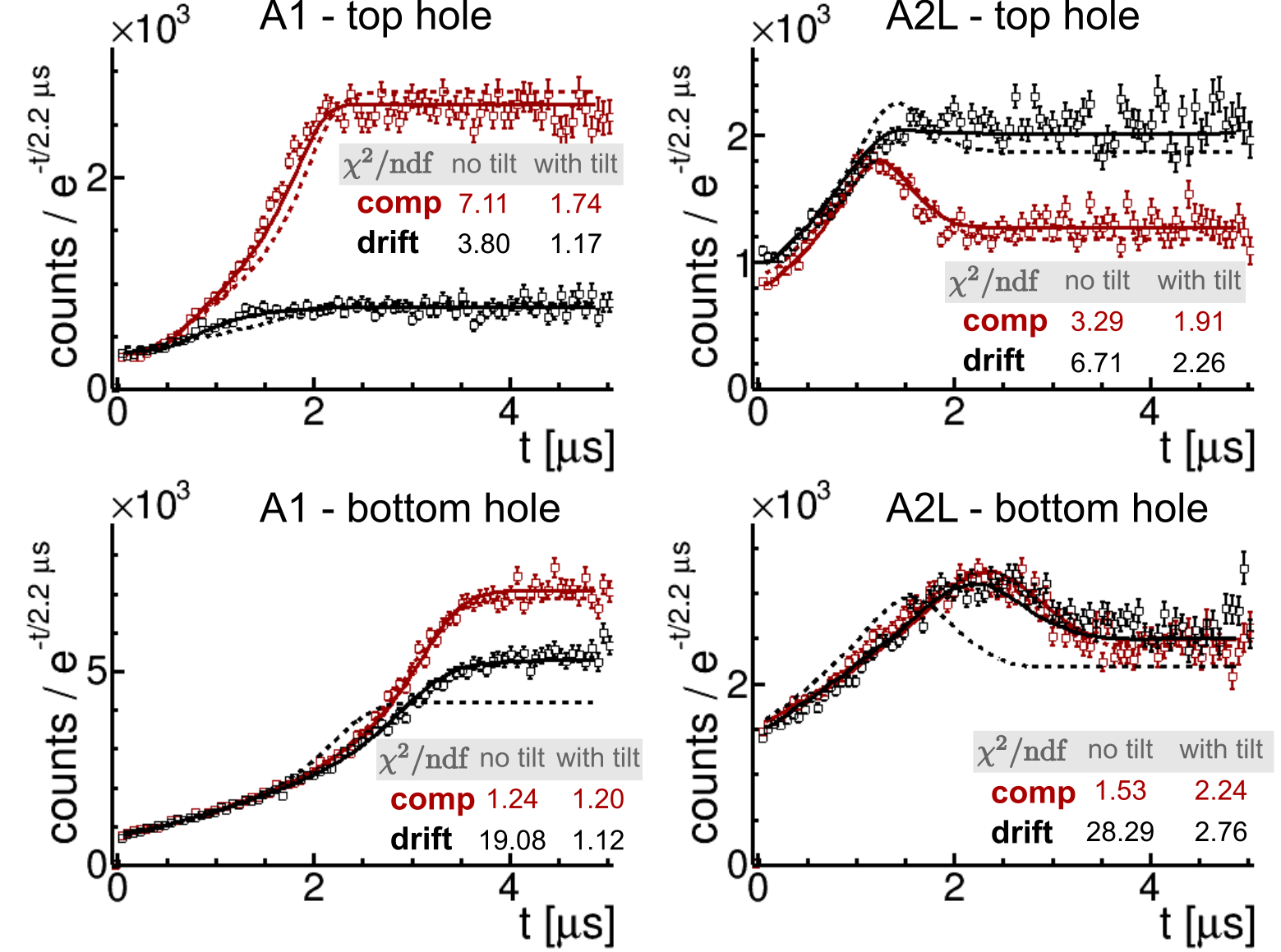}
	}
	
% \theaspect~words
	\caption{Measured (points) A1 and A2L time spectra for ``top-'' and ``bottom-hole'' injection and for ``compression'' (red) and ``pure-drift'' (black) conditions from Fig.~\ref{fig:trajectories}, along with corresponding simulated time spectra with (solid curves) and without (dashed curves) a small angular target misalignment. The error bars are given by the statistical errors of the measured counts (before the lifetime compensation) scaled by $ e^{t/2.2~\si{\micro \s}} $. Each simulated time spectrum was fitted to the measured spectra using two fit parameters (normalization and flat background). Reduced chi-squareds of each fit are also indicated ($ \text{ndf}=75 $). \label{fig:time spectra}}
\end{figure}

The measured A1 and A2L time spectra under ``compression'' conditions are presented in Fig.~\ref{fig:time spectra} (red dots), for both injection types. The time spectra were corrected for muon decay by multiplying the counts by $\exp(t/2.2~\mathrm{\mu s})$. The A1 counts increase with time, for both injection positions, indicating muons were gradually moving towards the tip of the target, \ie, towards the A1 acceptance region (see Fig.~\ref{fig:trans_detector_acc}). The A2L counts first increase, then decrease with time, suggesting muons first entered, then exited the detector acceptance region. After a certain time, the number of counts stays constant in both detectors, implying that the muons reached the target walls. For ``bottom-hole'' injection, this happens up to 1500~ns later compared to ``top-hole'' injection. Indeed, muons injected through the ``bottom hole'' travel through a denser gas than for ``top-hole'' injection, leading to a slower drift. All these features are consistent with the simulations of Fig.~\ref{fig:trajectories} (left).

The ``pure-drift'' time spectra for ``top-hole'' injection (black points in Fig.~\ref{fig:time spectra},  top row) differ significantly from the ``compression'' time spectra (red points). The A1 counts remain low at all times, suggesting muons never reached the tip of the target. The A2L counts stay constant after reaching the maximum, implying muons did not ``fly by'' the detector, but they stopped in the target walls within the acceptance region of the detector. This is consistent with the simulations of Fig.~\ref{fig:trajectories} (right).
However, for ``bottom-hole'' injection, the measured ``pure-drift'' time spectra (black points in Fig.~\ref{fig:time spectra}, bottom row) are almost identical to the ``compression'' time spectra (red points). This is mainly because of the worse detector resolution in the lower target region, which is further away from the detectors.

To better compare the measurements with the Geant4 simulations, the muon trajectories of Fig.~\ref{fig:trajectories} were folded with the detector acceptance of Fig.~\ref{fig:trans_detector_acc} to produce the corresponding time spectra.
%to obtain the time spectra of Fig.~\ref{fig:trans_newtop_comp} (red and black lines, for "compression" and "pure drift", respectively). 
These time spectra (dashed curves in Fig.~\ref{fig:time spectra}) were fitted to the measurements using two fit parameters per time spectrum: a normalization, accounting for uncertainties in the stopping probability and detection efficiencies (detectors were not calibrated), and a flat background, accounting for beam-related stops in the walls.
Additionally, the detector positions had to be varied in the simulations because of their position uncertainty (about 1 mm). The obtained chi-squareds are relatively large, which points to systematic discrepancies between the simulations and the measurements. 

We investigated the sensitivity of the simulated time spectra to other experimental uncertainties: initial beam characteristics, foil thicknesses,  electric and magnetic field maps, He density distribution, and misalignment (tilt) between the target and the magnetic field axes.
% 
% These sensitivity studies shows that the misalignment uncertainty by far dominates the uncertainty of the time spectra: the few mm position resolution of the downstream scintillators used to align the beam does not allow sufficiently precise knowledge of the muon stopping distribution. Note that even a 0.5\,mm displacement of the muon stopping distribution significantly affects the time spectra. 
% 
% These sensitivity studies shows that the alignment uncertainty which translates into a uncertainty of the muon stopping position by far dominates the uncertainty of the time spectra. Indeed already 0.5\,mm displacement of the muon stopping distribution significantly affects the time spectra. But such precise beam alignment is not possible with our downstream scintillators, which have position resolution of a few mm. 
% 
These sensitivity studies show that the alignment uncertainty, which translates into an uncertainty in muon stopping position, is by far the dominant one: even a 0.5\,mm muon-beam displacement significantly affects the time spectra. However, such precise beam alignment is not possible with our back detector (consisting of $4\times 4\times 5$~mm$^3$ scintillators). 
%detecting the positron spiriling around the B-field
%which has position resolution of a few mm.

% the few mm position resolution of the downstream scintillators used to align the beam does not allow sufficiently precise knowledge of the muon stopping distribution. Note that even a 0.5\,mm displacement of the muon stopping distribution significantly affects the time spectra. 

% Likely explanation of this discrepancy is a misalignment (tilt) between the target and the magnetic field axes because the few mm position resolution of the downstream scintillators used to align the beam, do not allow sufficient precise knowledge of the initial stopping position of the muon.
% Indeed, a sensitivity study shows that even a 0.5\,mm displacement of the muon stopping distribution significantly affects the time spectra. 
By simulating the time spectra for various tilts and fitting them to the data, we found that the best fit (for ``compression'' and ``pure drift'' simultaneously) is obtained by rotating the target axis from the $(0,0,1)$ direction to $(0.007,0.018,0.9998)$ for ``top-hole,'' and to $(0.019, 0.005, 0.9998)$ for ``bottom-hole'' injection. These tilts shift the center of the muon stopping distribution in the $ xy $-plane by up to $ \sim 3.5 $\,\si{\mm}. Note that the two different tilts are justified because the beam injection position was changed by adjusting the solenoid placement. 

Simulations including such tilts (solid curves in Fig.~\ref{fig:time spectra}) agree with measurements significantly better. Additional smaller improvements could be achieved by simultaneously fine-tuning all other experimental parameters (within their experimental uncertainty), together with tilt and detector positions.
However, little further insight would be gained from such a time-consuming optimization, since the simulations presented here already reproduce well the main features of the measurements. Hence, the measurements demonstrate the muon-beam transverse compression.

Next, we investigate the muon drift versus its energy by varying the electric field strength.
The drift angle $ \theta $ is proportional to the $\mu^+$-$\text{He}$ collision frequency given by
\begin{equation}
\nu_{c}=N\sigma_{MT}(E_{CM})\left|\vec{v}_{r}\right|,
\end{equation}
where $N$ is the helium number density, $\sigma_{MT}(E_{CM})$
the $\mu^+$-$\text{He}$ momentum transfer cross section, $E_{CM}$ the center-of-mass energy, and $\vec{v}_{r}$ the $\mu^+$-$\text{He}$ relative velocity.
\begin{figure}[]
		\testaspectone{
			\includegraphics[width=1\columnwidth]{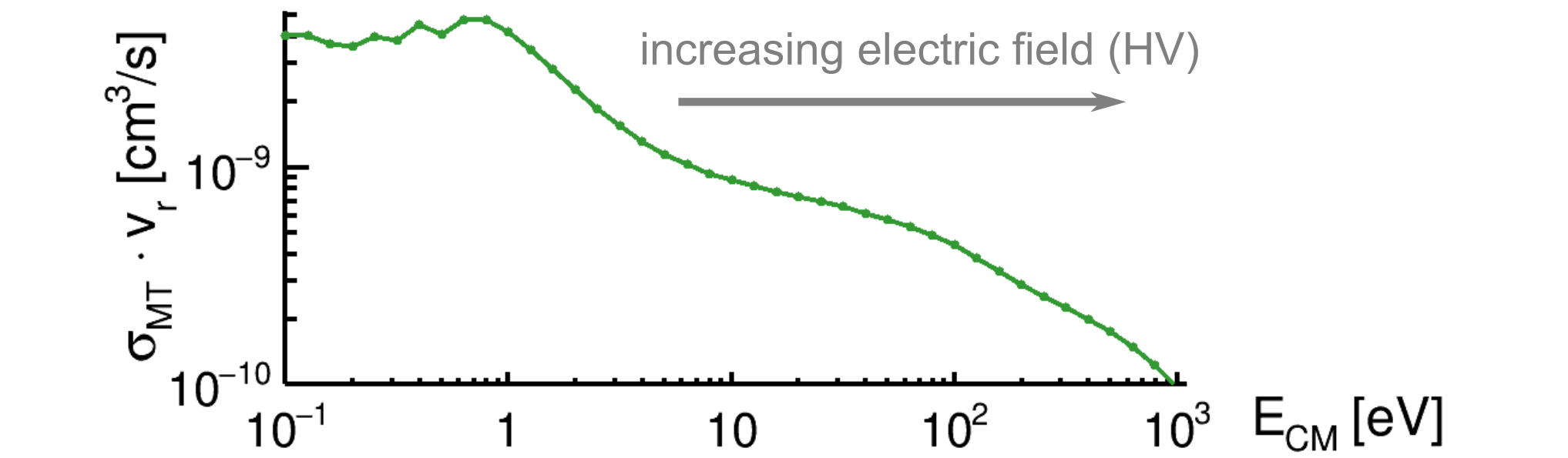}
		}
	
% \theaspect~words
	\caption{Product $\sigma_{MT} \cdot v_r$ vs. muon c.m. energy, using energy-scaled proton-He cross sections of Ref.~\cite{Krstic1998}. \label{fig:cs*sqrtE vs E}}
\end{figure}

\begin{figure}[]

			\testaspectone{	
		\includegraphics[width=1.0\columnwidth]{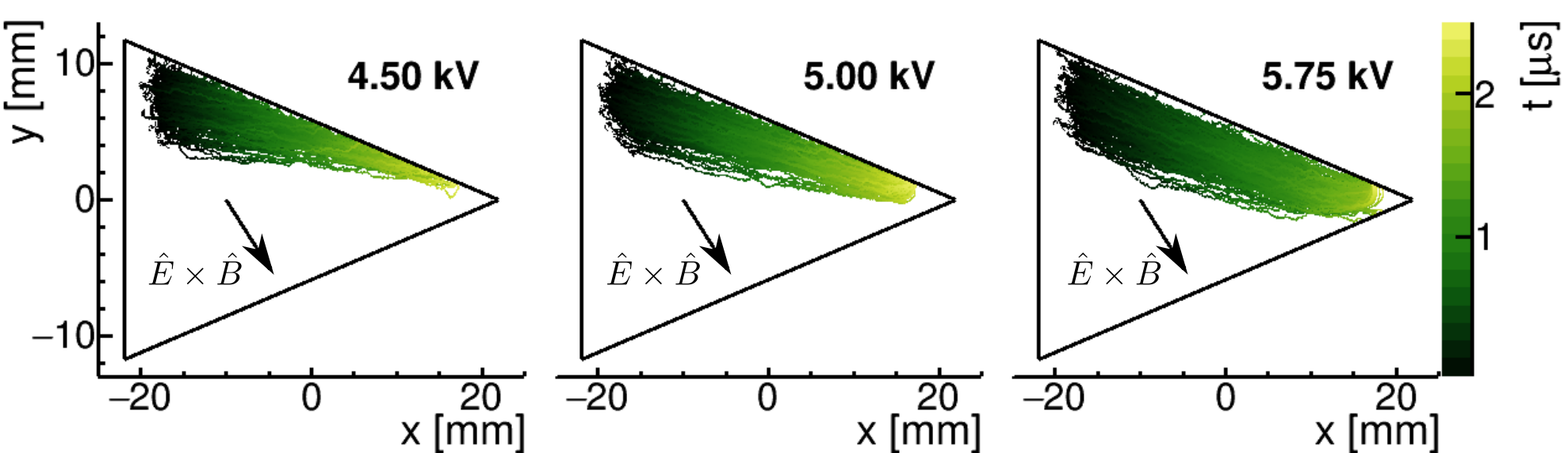}
	}

% \theaspect~words
	\caption{Muon trajectories in $ xy $-plane for  ``top-hole'' and ``compression'' conditions ($ 6 $ -- $19$~K at $ 8.6 $~mbar), for various HV. The target axis was assumed to point in the $(0.007,0.018,0.9998)$ direction, as determined previously by minimizing the chi-squared for the 5.0~kV measurements. \label{fig:trans_newtop_comp_sim_scan}}
\end{figure}

\begin{figure}[]

	\testaspectone{
		\includegraphics[width=1.0\columnwidth]{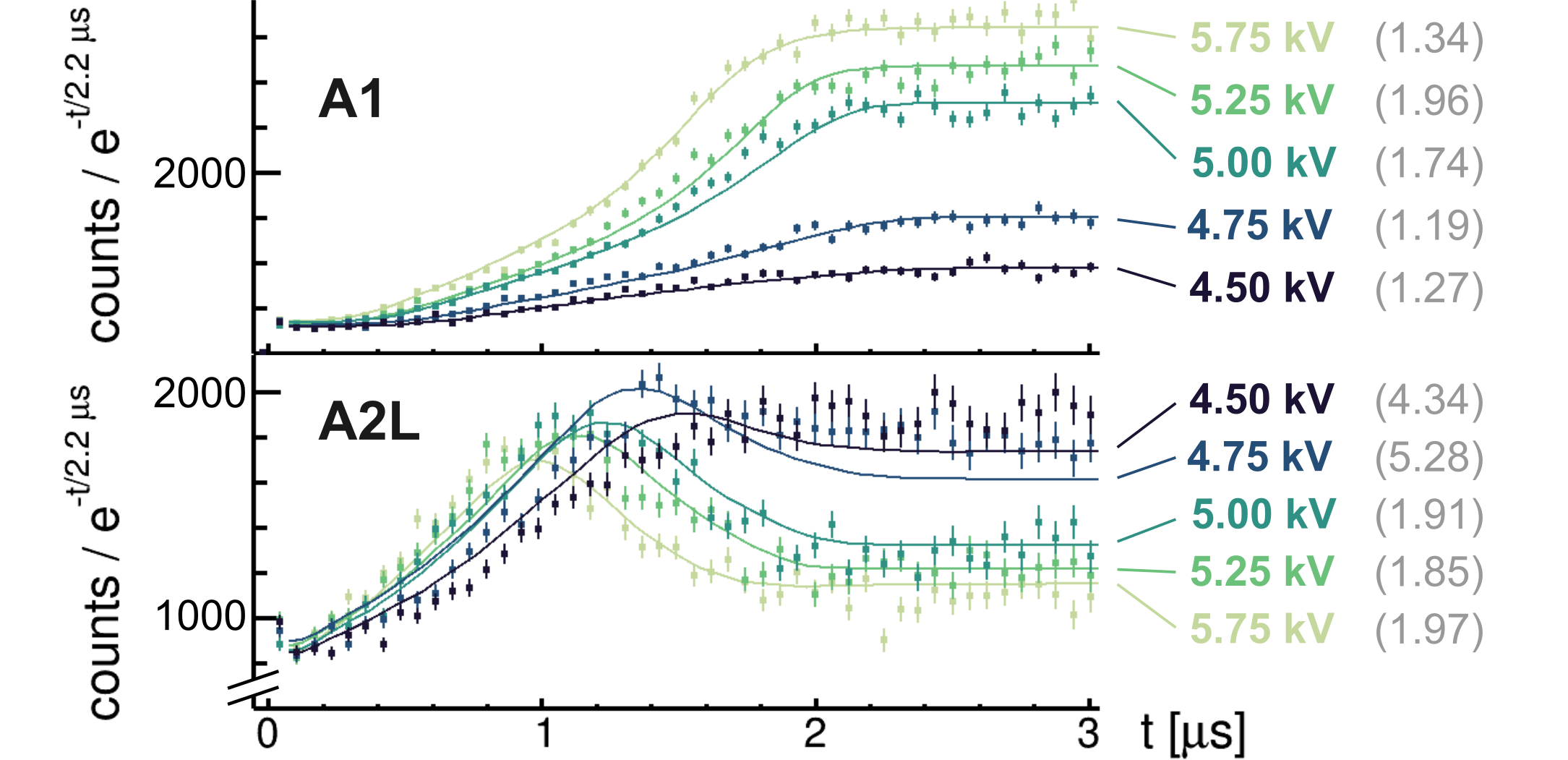}
	}
	
% \theaspect~words

	\caption{Measured (points) A1 and A2L time spectra for ``top-hole'' and ``compression'' configuration and for various HV, scaled to have the same number of injected muons. The error bars are purely statistical, as in Fig.~\ref{fig:time spectra}. The simulated time spectra (curves), corresponding to the trajectories of Fig.~\ref{fig:trans_newtop_comp_sim_scan}, were fitted to corresponding measured time spectra using two free parameters per spectrum (normalization and flat background). Chi-squareds are indicated in parentheses ($ \text{ndf}=75 $). \label{fig:trans_newtop_comp_data_scan}}
	%\end{figure}
\end{figure}

For muon energies $ \lesssim 1$~eV, $\sigma_{MT}$ scales as $1/\sqrt{E_{CM}}\propto 1/\left|\vec{v}_{r}\right|$~\cite{Mason1979b}, so that the collision frequency is independent of the muon energy. 
%However, this is only valid below $ \approx $1~eV.
Above $ 1 $~eV, the product $\sigma_{MT}\left|\vec{v}_{r}\right|$ decreases with energy (see Fig.~\ref{fig:cs*sqrtE vs E}), so that $\nu_c$ and $\theta$ decrease with increasing muon energy.
Hence, the muon drift direction approaches the \ExB-direction with increasing electric field (increasing energy), as shown in Fig.~\ref{fig:trans_newtop_comp_sim_scan}.
% Since the muon energy in the He gas increases with electric field strength, the muon drift direction approaches the \ExB-direction with increasing electric field. 
% The simulated muon trajectories of Fig.~\ref{fig:trans_newtop_comp_sim_scan} for various HV confirm this behavior.

The dependence of the muon motion on the electric field strength was measured for various experimental configurations. Figure~\ref{fig:trans_newtop_comp_data_scan} presents only the measurements for ``top-hole'' and ``compression'' configuration, which illustrate all relevant features: with decreasing HV, the A1 spectrum maxima decrease, indicating increasingly fewer muons reach the target tip, while the A2L spectra maxima shift towards later times, implying slower drift.
% Moreover, for smaller HV the drift is slower, as visible from the shift of the A2L time spectrum maximum towards later times. 
This is consistent with the simulated trajectories of Fig.~\ref{fig:trans_newtop_comp_sim_scan}.

The simulated time spectra for each HV and for each detector were fitted independently to the measurements with two free parameters: normalization and flat background.
A simultaneous fit for all HV values was not possible, as the normalization giving the best fit depended systematically on the HV. This might be caused by the lack of a precise definition of the electric field at the target tip, which would affect most significantly the measurements with strong electric fields, where muons reach the tip of the target.
A mismatch of the muon stopping distribution between simulations and experiment would also contribute to this systematic effect.

Still, the simulation correctly reproduces the muon drift velocity, which is given by the energy-dependent $\mu^+$-$\text{He}$ elastic cross section. Indeed, the observed maxima of the time spectra, occurring when muons arrive in the detector acceptance region, are consistent with simulations.

To summarize, in this Letter we demonstrate muon-beam phase-space compression in the $y$-direction with the muCool device. One critical aspect of this demonstration was distinguishing between simple muon drift versus drift with simultaneous reduction of the beam vertical size. Such distinction was accomplished by performing measurements with and without a vertical temperature gradient, which is needed for transverse compression, and by injecting the muon beam into different vertical target regions (up to 14\, mm apart in the $y$-direction as shown in Fig.~\ref{fig:trajectories}, left). 
% By combining the measurements with different beam injections, we demonstrated that we are able to compress muons that are initially 14 mm apart to a 0.25\,mm (RMS) ?spot? (see Fig.~\ref{fig:trajectories} (left)).
The muon motion was simulated using the Geant4 toolkit including custom low-energy ($ <1 $ keV) muon processes~\cite{Belosevic2020}. Very good agreement with the measurements was achieved after accounting for small target and detector misalignments. Our modeling of low-energy $\mu^+$-$\text{He}$ collisions was further validated by studying the dependence of the muon drift on the applied electric field.

% According to the simulations, a muon beam of 10~mm diameter and 830~keV energy with 17~keV spread and 600 keV/c momentum spread (in each direction) is transformed within 3.5~\si{\micro \s} into a beam of 0.7~mm size (FWHM) in the $ y $-direction and 5~eV energy with 10~keV momentum spread (in all directions).
Simulations show that after the transverse compression stage the muon beam has $\sigma_y=0.25$~mm, 5~eV energy, and $\sigma_{p_x,p_y,p_z} \approx 20$~keV/c.
These beam characteristics depend only on the electric field strength and gas density (gradient) regardless of the initial beam properties.
When applying this transverse stage to a typical standard high intensity muon beam (e.g. $\pi$E5~\cite{Adam2013,baldini2013meg,PSI_muE5_web} and $\mu$E4~\cite{Prokscha2008} surface beams at PSI) with  $\sigma_{x,y}\approx 10$~mm  and $\sigma_{ p_x,p_y} \approx2$~MeV/c at 28 MeV/c momentum (4 MeV energy), we obtain a $\sigma_y\cdot\sigma_{p_y}$ phase-space compression by a factor of $\sim 4\cdot 10^3 $ with an overall efficiency of 32\%. This efficiency is mainly given by muon decay, as the compression takes on average $2.3$~\si{\micro \s}, and includes 10\% losses from muonium formation. We assumed here 100\% stopping probability in He gas (achievable by slightly elongating the target in the $z$-direction).
% The final beam phase space density is largely independent of the initial beam parameters (as long as the beam can be efficiently stopped in the transverse target).

% When the transverse compression stage demonstrated in this Letter is connected to the previously demonstrated longitudinal compression stage~\cite{Bao2014,Belosevic2019} and extracted
While $\sigma_{p_x}$ and $\sigma_{p_z}$ are reduced by the dissipative slowing-down process in the transverse stage, the extension in the $z$-direction $\sigma_z$ is reduced only in the following longitudinal stage, while $\sigma_x$ is reduced by a pulsed re-acceleration process after muon extraction~\cite{Taqqu2006}.
Therefore, a total phase-space compression of $ 10^{10} $ with $ 10^{-3} $ efficiency is attainable by connecting the transverse compression stage demonstrated in this Letter to the previously demonstrated longitudinal compression stage~\cite{Bao2014,Belosevic2019}, followed by vacuum-extraction and re-acceleration~\cite{Taqqu2006}.

\begin{acknowledgements}
The experimental work was performed at the $ \pi E 1 $ beamline at the PSI proton accelerator HIPA. We thank the machine and beam line groups for providing excellent conditions. Simulations were performed on the Euler cluster managed by the HPC team at ETH Z\"urich.
We gratefully acknowledge the outstanding support received from the workshops and support groups at ETH Zurich and PSI. Furthermore, we thank F. Kottmann, M. Horisberger, U. Greuter, R. Scheuermann, T. Prokscha, D. Reggiani, K. Deiters, T. Rauber, and F. Barchetti for their help. This work was supported by the SNF grants No. 200020\_159754 and 200020\_172639.
\end{acknowledgements}

%\bibliography{library}
% \bibliography{ref_prl}

\begin{thebibliography}{20}%
	\makeatletter
	\providecommand \@ifxundefined [1]{%
		\@ifx{#1\undefined}
	}%
	\providecommand \@ifnum [1]{%
		\ifnum #1\expandafter \@firstoftwo
		\else \expandafter \@secondoftwo
		\fi
	}%
	\providecommand \@ifx [1]{%
		\ifx #1\expandafter \@firstoftwo
		\else \expandafter \@secondoftwo
		\fi
	}%
	\providecommand \natexlab [1]{#1}%
	\providecommand \enquote [1]{``#1''}%
	\providecommand \bibnamefont [1]{#1}%
	\providecommand \bibfnamefont [1]{#1}%
	\providecommand \citenamefont [1]{#1}%
	\providecommand \href@noop [0]{\@secondoftwo}%
	\providecommand \href [0]{\begingroup \@sanitize@url \@href}%
	\providecommand \@href[1]{\@@startlink{#1}\@@href}%
	\providecommand \@@href[1]{\endgroup#1\@@endlink}%
	\providecommand \@sanitize@url [0]{\catcode `\\12\catcode `\$12\catcode
		`\&12\catcode `\#12\catcode `\^12\catcode `\_12\catcode `\%12\relax}%
	\providecommand \@@startlink[1]{}%
	\providecommand \@@endlink[0]{}%
	\providecommand \url [0]{\begingroup\@sanitize@url \@url }%
	\providecommand \@url [1]{\endgroup\@href {#1}{\urlprefix }}%
	\providecommand \urlprefix [0]{URL }%
	\providecommand \Eprint [0]{\href }%
	\providecommand \doibase [0]{https://doi.org/}%
	\providecommand \selectlanguage [0]{\@gobble}%
	\providecommand \bibinfo [0]{\@secondoftwo}%
	\providecommand \bibfield [0]{\@secondoftwo}%
	\providecommand \translation [1]{[#1]}%
	\providecommand \BibitemOpen [0]{}%
	\providecommand \bibitemStop [0]{}%
	\providecommand \bibitemNoStop [0]{.\EOS\space}%
	\providecommand \EOS [0]{\spacefactor3000\relax}%
	\providecommand \BibitemShut [1]{\csname bibitem#1\endcsname}%
	\let\auto@bib@innerbib\@empty
	%</preamble>
	\bibitem [{\citenamefont {Gorringe}\ and\ \citenamefont
		{Hertzog}(2015)}]{Gorringe2015}%
	\BibitemOpen
	\bibfield {author} {\bibinfo {author} {\bibfnamefont {T.}~\bibnamefont
			{Gorringe}}\ and\ \bibinfo {author} {\bibfnamefont {D.}~\bibnamefont
			{Hertzog}},\ }\href {https://doi.org/10.1016/j.ppnp.2015.06.001} {\bibfield
		{journal} {\bibinfo {journal} {Prog. Part. Nucl. Phys.}\ }\textbf {\bibinfo
			{volume} {84}},\ \bibinfo {pages} {73} (\bibinfo {year} {2015})}\BibitemShut
	{NoStop}%
	\bibitem [{\citenamefont {Iinuma}(2011)}]{Iinuma2011}%
	\BibitemOpen
	\bibfield {author} {\bibinfo {author} {\bibfnamefont {H.}~\bibnamefont
			{Iinuma}},\ }\href {https://doi.org/10.1088/1742-6596/295/1/012032}
	{\bibfield {journal} {\bibinfo {journal} {J. Phys. Conf. Ser.}\ }\textbf
		{\bibinfo {volume} {295}},\ \bibinfo {pages} {012032} (\bibinfo {year}
		{2011})}\BibitemShut {NoStop}%
	\bibitem [{\citenamefont {Adelmann}\ \emph {et~al.}(2010)\citenamefont
		{Adelmann}, \citenamefont {Kirch}, \citenamefont {Onderwater},\ and\
		\citenamefont {Schietinger}}]{Adelmann2010}%
	\BibitemOpen
	\bibfield {author} {\bibinfo {author} {\bibfnamefont {A.}~\bibnamefont
			{Adelmann}}, \bibinfo {author} {\bibfnamefont {K.}~\bibnamefont {Kirch}},
		\bibinfo {author} {\bibfnamefont {C.~J.~G.}\ \bibnamefont {Onderwater}},\
		and\ \bibinfo {author} {\bibfnamefont {T.}~\bibnamefont {Schietinger}},\
	}\href {https://doi.org/10.1088/0954-3899/37/8/085001} {\bibfield {journal}
		{\bibinfo {journal} {J. Phys. G}\ }\textbf {\bibinfo {volume} {37}},\
		\bibinfo {pages} {085001} (\bibinfo {year} {2010})}\BibitemShut {NoStop}%
	\bibitem [{\citenamefont {Crivellin}\ \emph {et~al.}(2018)\citenamefont
		{Crivellin}, \citenamefont {Hoferichter},\ and\ \citenamefont
		{Schmidt-Wellenburg}}]{Crivellin2018}%
	\BibitemOpen
	\bibfield {author} {\bibinfo {author} {\bibfnamefont {A.}~\bibnamefont
			{Crivellin}}, \bibinfo {author} {\bibfnamefont {M.}~\bibnamefont
			{Hoferichter}},\ and\ \bibinfo {author} {\bibfnamefont {P.}~\bibnamefont
			{Schmidt-Wellenburg}},\ }\href {https://doi.org/10.1103/PhysRevD.98.113002}
	{\bibfield {journal} {\bibinfo {journal} {Phys. Rev. D}\ }\textbf {\bibinfo
			{volume} {98}},\ \bibinfo {pages} {113002} (\bibinfo {year}
		{2018})}\BibitemShut {NoStop}%
	\bibitem [{\citenamefont {Crivelli}(2018)}]{Cr}%
	\BibitemOpen
	\bibfield {author} {\bibinfo {author} {\bibfnamefont {P.}~\bibnamefont
			{Crivelli}},\ }\href {https://doi.org/10.1007/s10751-018-1525-z} {\bibfield
		{journal} {\bibinfo {journal} {Hyperfine Interact.}\ }\textbf {\bibinfo
			{volume} {239}},\ \bibinfo {pages} {49} (\bibinfo {year} {2018})}\BibitemShut
	{NoStop}%
	\bibitem [{\citenamefont {Kirch}\ and\ \citenamefont
		{Khaw}(2014)}]{Kirch2014a}%
	\BibitemOpen
	\bibfield {author} {\bibinfo {author} {\bibfnamefont {K.}~\bibnamefont
			{Kirch}}\ and\ \bibinfo {author} {\bibfnamefont {K.~S.}\ \bibnamefont
			{Khaw}},\ }\href {https://doi.org/10.1142/S2010194514602580} {\bibfield
		{journal} {\bibinfo {journal} {Int. J. Mod. Phys. Conf. Ser.}\ }\textbf
		{\bibinfo {volume} {30}},\ \bibinfo {pages} {1460258} (\bibinfo {year}
		{2014})}\BibitemShut {NoStop}%
	\bibitem [{\citenamefont {Antognini}\ \emph {et~al.}(2018)\citenamefont
		{Antognini} \emph {et~al.}}]{Kaplan2018}%
	\BibitemOpen
	\bibfield {author} {\bibinfo {author} {\bibfnamefont {A.}~\bibnamefont
			{Antognini}} \emph {et~al.},\ }\href {https://doi.org/10.3390/atoms6020017}
	{\bibfield {journal} {\bibinfo {journal} {Atoms}\ }\textbf {\bibinfo
			{volume} {6}},\ \bibinfo {pages} {17} (\bibinfo {year} {2018})}\BibitemShut
	{NoStop}%
	\bibitem [{\citenamefont {van~der Meer}(1985)}]{VanderMeer1985}%
	\BibitemOpen
	\bibfield {author} {\bibinfo {author} {\bibfnamefont {S.}~\bibnamefont
			{van~der Meer}},\ }\href {https://doi.org/10.1103/RevModPhys.57.689}
	{\bibfield {journal} {\bibinfo {journal} {Rev. Mod. Phys.}\ }\textbf
		{\bibinfo {volume} {57}},\ \bibinfo {pages} {689} (\bibinfo {year}
		{1985})}\BibitemShut {NoStop}%
	\bibitem [{\citenamefont {Budker}\ and\ \citenamefont
		{Skrinskiĭ}(1978)}]{Budker1978}%
	\BibitemOpen
	\bibfield {author} {\bibinfo {author} {\bibfnamefont {G.~I.}\ \bibnamefont
			{Budker}}\ and\ \bibinfo {author} {\bibfnamefont {A.~N.}\ \bibnamefont
			{Skrinskiĭ}},\ }\href {https://doi.org/10.1070/PU1978v021n04ABEH005537}
	{\bibfield {journal} {\bibinfo {journal} {Sov. Phys. Uspekhi}\ }\textbf
		{\bibinfo {volume} {21}},\ \bibinfo {pages} {277} (\bibinfo {year}
		{1978})}\BibitemShut {NoStop}%
	\bibitem [{\citenamefont {M{\"{u}}hlbauer}\ \emph {et~al.}(1999)\citenamefont
		{M{\"{u}}hlbauer}, \citenamefont {Daniel}, \citenamefont {Hartmann},
		\citenamefont {Hauser}, \citenamefont {Kottmann}, \citenamefont {Petitjean},
		\citenamefont {Schott}, \citenamefont {Taqqu},\ and\ \citenamefont
		{Wojciechowski}}]{Muhlbauer1999}%
	\BibitemOpen
	\bibfield {author} {\bibinfo {author} {\bibfnamefont {M.}~\bibnamefont
			{M{\"{u}}hlbauer}}, \bibinfo {author} {\bibfnamefont {H.}~\bibnamefont
			{Daniel}}, \bibinfo {author} {\bibfnamefont {F.}~\bibnamefont {Hartmann}},
		\bibinfo {author} {\bibfnamefont {P.}~\bibnamefont {Hauser}}, \bibinfo
		{author} {\bibfnamefont {F.}~\bibnamefont {Kottmann}}, \bibinfo {author}
		{\bibfnamefont {C.}~\bibnamefont {Petitjean}}, \bibinfo {author}
		{\bibfnamefont {W.}~\bibnamefont {Schott}}, \bibinfo {author} {\bibfnamefont
			{D.}~\bibnamefont {Taqqu}},\ and\ \bibinfo {author} {\bibfnamefont
			{P.}~\bibnamefont {Wojciechowski}},\ }\href
	{https://doi.org/10.1023/A:1012624501134} {\bibfield {journal} {\bibinfo
			{journal} {Hyperfine Interact.}\ }\textbf {\bibinfo {volume} {119}},\
		\bibinfo {pages} {305} (\bibinfo {year} {1999})}\BibitemShut {NoStop}%
	\bibitem [{\citenamefont {Morenzoni}\ \emph {et~al.}(1994)\citenamefont
		{Morenzoni}, \citenamefont {Kottmann}, \citenamefont {Maden}, \citenamefont
		{Matthias}, \citenamefont {Meyberg}, \citenamefont {Prokscha}, \citenamefont
		{Wutzke},\ and\ \citenamefont {Zimmermann}}]{Morenzoni1994}%
	\BibitemOpen
	\bibfield {author} {\bibinfo {author} {\bibfnamefont {E.}~\bibnamefont
			{Morenzoni}}, \bibinfo {author} {\bibfnamefont {F.}~\bibnamefont {Kottmann}},
		\bibinfo {author} {\bibfnamefont {D.}~\bibnamefont {Maden}}, \bibinfo
		{author} {\bibfnamefont {B.}~\bibnamefont {Matthias}}, \bibinfo {author}
		{\bibfnamefont {M.}~\bibnamefont {Meyberg}}, \bibinfo {author} {\bibfnamefont
			{T.}~\bibnamefont {Prokscha}}, \bibinfo {author} {\bibfnamefont
			{T.}~\bibnamefont {Wutzke}},\ and\ \bibinfo {author} {\bibfnamefont
			{U.}~\bibnamefont {Zimmermann}},\ }\href
	{https://doi.org/10.1103/PhysRevLett.72.2793} {\bibfield {journal} {\bibinfo
			{journal} {Phys. Rev. Lett.}\ }\textbf {\bibinfo {volume} {72}},\ \bibinfo
		{pages} {2793} (\bibinfo {year} {1994})}\BibitemShut {NoStop}%
\bibitem [{\citenamefont {Pifer}\ \emph {et~al.}(1976)\citenamefont {Pifer},
	\citenamefont {Bowen},\ and\ \citenamefont {Kendall}}]{Pifer1976a}%
\BibitemOpen
\bibfield  {author} {\bibinfo {author} {\bibfnamefont {A.~E.}\ \bibnamefont
		{Pifer}}, \bibinfo {author} {\bibfnamefont {T.}~\bibnamefont {Bowen}},\ and\
	\bibinfo {author} {\bibfnamefont {K.~R.}\ \bibnamefont {Kendall}},\ }\href
{https://doi.org/10.1016/0029-554X(76)90823-5} {\bibfield  {journal}
	{\bibinfo  {journal} {Nucl. Instr. and Meth.}\ }\textbf {\bibinfo
		{volume} {135}},\ \bibinfo {pages} {39} (\bibinfo {year} {1976})}\BibitemShut
{NoStop}%
	\bibitem [{\citenamefont {Taqqu}(2006)}]{Taqqu2006}%
	\BibitemOpen
	\bibfield {author} {\bibinfo {author} {\bibfnamefont {D.}~\bibnamefont
			{Taqqu}},\ }\href {https://doi.org/10.1103/PhysRevLett.97.194801} {\bibfield
		{journal} {\bibinfo {journal} {Phys. Rev. Lett.}\ }\textbf {\bibinfo
			{volume} {97}},\ \bibinfo {pages} {194801} (\bibinfo {year}
		{2006})}\BibitemShut {NoStop}%
	\bibitem [{\citenamefont {Heylen}(1980)}]{Heylen1980}%
	\BibitemOpen
	\bibfield {author} {\bibinfo {author} {\bibfnamefont {A.~E.~D.}\
			\bibnamefont {Heylen}},\ }\href
	{https://ieeexplore.ieee.org/abstract/document/4648132} {\bibfield {journal}
		{\bibinfo {journal} {IEE Proc. A}\ }\textbf {\bibinfo {volume} {127}},\
		\bibinfo {pages} {221} (\bibinfo {year} {1980})}\BibitemShut {NoStop}%
	\bibitem [{\citenamefont {Bao}\ \emph {et~al.}(2014)\citenamefont {Bao} \emph
		{et~al.}}]{Bao2014}%
	\BibitemOpen
	\bibfield {author} {\bibinfo {author} {\bibfnamefont {Y.}~\bibnamefont
			{Bao}} \emph {et~al.} (\bibinfo {collaboration} {muCool collaboration}),\
	}\href {https://doi.org/10.1103/PhysRevLett.112.224801} {\bibfield {journal}
		{\bibinfo {journal} {Phys. Rev. Lett.}\ }\textbf {\bibinfo {volume} {112}},\
		\bibinfo {pages} {224801} (\bibinfo {year} {2014})}\BibitemShut {NoStop}%
	\bibitem [{\citenamefont {Belosevic}\ \emph {et~al.}(2019)\citenamefont
		{Belosevic} \emph {et~al.}}]{Belosevic2019}%
	\BibitemOpen
	\bibfield {author} {\bibinfo {author} {\bibfnamefont {I.}~\bibnamefont
			{Belosevic}} \emph {et~al.} (\bibinfo {collaboration} {muCool
			collaboration}),\ }\href {https://doi.org/10.1140/epjc/s10052-019-6932-z}
	{\bibfield {journal} {\bibinfo {journal} {Eur. Phys. J. C}\ }\textbf
		{\bibinfo {volume} {79}},\ \bibinfo {pages} {430} (\bibinfo {year}
		{2019})}\BibitemShut {NoStop}%
	\bibitem [{\citenamefont {Wichmann}\ \emph {et~al.}(2016)\citenamefont
		{Wichmann}, \citenamefont {Antognini}, \citenamefont {Eggenberger},
		\citenamefont {Kirch}, \citenamefont {Piegsa}, \citenamefont {Soler},
		\citenamefont {Stahn},\ and\ \citenamefont {Taqqu}}]{Wichmann2016a}%
	\BibitemOpen
	\bibfield {author} {\bibinfo {author} {\bibfnamefont {G.}~\bibnamefont
			{Wichmann}}, \bibinfo {author} {\bibfnamefont {A.}~\bibnamefont {Antognini}},
		\bibinfo {author} {\bibfnamefont {A.}~\bibnamefont {Eggenberger}}, \bibinfo
		{author} {\bibfnamefont {K.}~\bibnamefont {Kirch}}, \bibinfo {author}
		{\bibfnamefont {F.~M.}\ \bibnamefont {Piegsa}}, \bibinfo {author}
		{\bibfnamefont {U.}~\bibnamefont {Soler}}, \bibinfo {author} {\bibfnamefont
			{J.}~\bibnamefont {Stahn}},\ and\ \bibinfo {author} {\bibfnamefont
			{D.}~\bibnamefont {Taqqu}},\ }\href
	{https://doi.org/10.1016/j.nima.2016.01.018} {\bibfield {journal} {\bibinfo
			{journal} {Nucl. Instruments Methods Phys. Res. Sect. A Accel. Spectrometers,
				Detect. Assoc. Equip.}\ }\textbf {\bibinfo {volume} {814}},\ \bibinfo {pages}
		{33} (\bibinfo {year} {2016})}\BibitemShut {NoStop}%
	\bibitem [{\citenamefont {Agostinelli}\ \emph {et~al.}(2003)\citenamefont
		{Agostinelli} \emph {et~al.}}]{Agostinelli2003}%
	\BibitemOpen
	\bibfield {author} {\bibinfo {author} {\bibfnamefont {S.}~\bibnamefont
			{Agostinelli}} \emph {et~al.} (\bibinfo {collaboration} {Geant4
			collaboration}),\ }\href {https://doi.org/10.1016/S0168-9002(03)01368-8}
	{\bibfield {journal} {\bibinfo {journal} {Nucl. Instruments Methods Phys.
				Res. Sect. A Accel. Spectrometers, Detect. Assoc. Equip.}\ }\textbf {\bibinfo
			{volume} {506}},\ \bibinfo {pages} {250} (\bibinfo {year}
		{2003})}\BibitemShut {NoStop}%
	\bibitem [{\citenamefont {Belosevic}(2019)}]{Belosevic2020}%
	\BibitemOpen
	\bibfield {author} {\bibinfo {author} {\bibfnamefont {I.}~\bibnamefont
			{Belosevic}},\ }\emph {\bibinfo {title} {{Simulation and experimental
				verification of transverse and longitudinal compression of positive muon
				beams: Towards a novel high-brightness low-energy muon beam-line}}},\ \href
	{https://doi.org/10.3929/ETHZ-B-000402802} {Ph.D. thesis},\ \bibinfo
	{school} {ETH Zurich} (\bibinfo {year} {2019})\BibitemShut {NoStop}%
	\bibitem [{\citenamefont {Schultz}\ and\ \citenamefont
		{Krstic}(1998)}]{Krstic1998}%
	\BibitemOpen
	\bibfield {author} {\bibinfo {author} {\bibfnamefont {D.~R.}\ \bibnamefont
			{Schultz}}\ and\ \bibinfo {author} {\bibfnamefont {P.~S.}\ \bibnamefont
			{Krstic}},\ }\href {https://www-amdis.iaea.org/ALADDIN/} {\emph {\bibinfo
			{title} {At. plasma-material Interact. data fusion}}},\ Vol.~\bibinfo
	{volume} {8}\ (\bibinfo {year} {1998})\BibitemShut {NoStop}%
	\bibitem [{\citenamefont {Mason}\ \emph {et~al.}(1979)\citenamefont {Mason},
		\citenamefont {Lin},\ and\ \citenamefont {Gatland}}]{Mason1979b}%
	\BibitemOpen
	\bibfield {author} {\bibinfo {author} {\bibfnamefont {E.~A.}\ \bibnamefont
			{Mason}}, \bibinfo {author} {\bibfnamefont {S.~L.}\ \bibnamefont {Lin}},\
		and\ \bibinfo {author} {\bibfnamefont {I.~R.}\ \bibnamefont {Gatland}},\
	}\href {https://doi.org/10.1088/0022-3700/12/24/023} {\bibfield {journal}
		{\bibinfo {journal} {J. Phys. B At. Mol. Phys.}\ }\textbf {\bibinfo {volume}
			{12}},\ \bibinfo {pages} {4179} (\bibinfo {year} {1979})}\BibitemShut
	{NoStop}%
\bibitem [{\citenamefont {Adam}\ \emph {et~al.}(2013)\citenamefont {Adam},
	\citenamefont {Bai}, \citenamefont {Baldini}, \citenamefont {Baracchini},
	\citenamefont {Bemporad}, \citenamefont {Boca}, \citenamefont {Cattaneo},
	\citenamefont {Cavoto}, \citenamefont {Cei}, \citenamefont {Cerri} \emph
	{et~al.}}]{Adam2013}%
\BibitemOpen
\bibfield  {author} {\bibinfo {author} {Adam, J.}, \bibinfo {author} {Bai,
		X.}, \bibinfo {author} {Baldini, A.M.}, \bibinfo {author} {Baracchini, E.},
	\bibinfo {author} {Bemporad, C.}, \bibinfo {author} {Boca, G.}, \bibinfo
	{author} {Cattaneo, P.W.}, \bibinfo {author} {Cavoto, G.}, \bibinfo {author}
	{Cei, F.}, \bibinfo {author} {Cerri, C.}, et~al.,\ }\href
{https://doi.org/10.1140/epjc/s10052-013-2365-2} {\bibfield  {journal}
	{\bibinfo  {journal} {The European Physical Journal C}\ }\textbf {\bibinfo
		{volume} {73}},\ \bibinfo {pages} {2365} (\bibinfo {year}
	{2013})}\BibitemShut {NoStop}%
\bibitem [{\citenamefont {Baldini}\ \emph {et~al.}(2013)\citenamefont
	{Baldini}, \citenamefont {Cei}, \citenamefont {Cerri}, \citenamefont
	{Dussoni}, \citenamefont {Galli}, \citenamefont {Grassi}, \citenamefont
	{Nicol\`{o}}, \citenamefont {Raffaelli}, \citenamefont {Sergiampietri},
	\citenamefont {Signorelli}, \citenamefont {Tenchini}, \citenamefont
	{Bagliani}, \citenamefont {Gerone}, \citenamefont {Gatti}, \citenamefont
	{Baracchini}, \citenamefont {Fujii}, \citenamefont {Iwamoto}, \citenamefont
	{Kaneko}, \citenamefont {Mori}, \citenamefont {Nishimura}, \citenamefont
	{Ootani}, \citenamefont {Sawada}, \citenamefont {Uchiyama}, \citenamefont
	{Boca}, \citenamefont {Cattaneo}, \citenamefont {de~Bari}, \citenamefont
	{Nard�}, \citenamefont {Rossella}, \citenamefont {Cascella}, \citenamefont
	{Grancagnolo}, \citenamefont {L'Erario}, \citenamefont {Maffezzoli},
	\citenamefont {Miccoli}, \citenamefont {Onorato}, \citenamefont {Palam�},
	\citenamefont {Panareo}, \citenamefont {Pepino}, \citenamefont {Rella},
	\citenamefont {Tassielli}, \citenamefont {Zavarise}, \citenamefont {Cavoto},
	\citenamefont {Graziosi}, \citenamefont {Piredda}, \citenamefont {Ripiccini},
	\citenamefont {Voena}, \citenamefont {Grigoriev}, \citenamefont {Ignatov},
	\citenamefont {Khazin}, \citenamefont {Popov}, \citenamefont {Yudin},
	\citenamefont {Haruyama}, \citenamefont {Mihara}, \citenamefont {Nishiguchi},
	\citenamefont {Yamamoto}, \citenamefont {Hildebrandt}, \citenamefont
	{Kettle}, \citenamefont {Papa}, \citenamefont {Renga}, \citenamefont {Ritt},
	\citenamefont {Stoykov}, \citenamefont {Kang}, \citenamefont {Lim},
	\citenamefont {Molzon}, \citenamefont {You}, \citenamefont {Khomutov},
	\citenamefont {Korenchenko}, \citenamefont {Kravchuk},\ and\ \citenamefont
	{Kuchinksy}}]{baldini2013meg}%
\BibitemOpen
\bibfield  {author} {\bibinfo {author} {Baldini, A.M.}, \bibinfo {author}
	{Cei, F.}, \bibinfo {author} {Cerri, C.}, \bibinfo {author} {Dussoni, S.},
	\bibinfo {author} {Galli, L.}, \bibinfo {author} {Grassi, M.}, \bibinfo
	{author} {Nicol\`{o}, D.}, \bibinfo {author} {Raffaelli, F.}, \bibinfo
	{author} {Sergiampietri, F.}, \bibinfo {author} {Signorelli, G.}, et~al.,\
}\href@noop {} {\bibinfo {title} {Meg upgrade proposal}} (\bibinfo {year}
{2013}),\ \Eprint {https://arxiv.org/abs/1301.7225} {arXiv:1301.7225
	[physics.ins-det]} \BibitemShut {NoStop}%
\bibitem [{\citenamefont {{Paul Scherer Institute, Secondary beam lines at
			PSI}}()}]{PSI_muE5_web}%
\BibitemOpen
\bibfield  {author} {\bibinfo {author} {{Paul Scherer Institute, Secondary
			beam lines at PSI}},\ }\href {https://www.psi.ch/en/sbl/pie5-beamline}
{\bibinfo {title} {{https://www.psi.ch/en/sbl/pie5-beamline}}}\BibitemShut
{NoStop}%
\bibitem [{\citenamefont {Prokscha}\ \emph {et~al.}(2008)\citenamefont
	{Prokscha}, \citenamefont {Morenzoni}, \citenamefont {Deiters}, \citenamefont
	{Foroughi}, \citenamefont {George}, \citenamefont {Kobler}, \citenamefont
	{Suter},\ and\ \citenamefont {Vrankovic}}]{Prokscha2008}%
\BibitemOpen
\bibfield  {author} {\bibinfo {author} {Prokscha, T.}, \bibinfo {author}
	{Morenzoni, E.}, \bibinfo {author} {Deiters, K.}, \bibinfo {author}
	{Foroughi, F.}, \bibinfo {author} {George, D.}, \bibinfo {author} {Kobler,
		R.}, \bibinfo {author} {Suter, A.}, \bibinfo {author} {Vrankovic, V.},\
}\href {https://doi.org/10.1016/j.nima.2008.07.081} {\bibfield  {journal}
	{\bibinfo  {journal} {Nuclear Instruments and Methods in Physics Research,
			Section A: Accelerators, Spectrometers, Detectors and Associated Equipment}\
	}\textbf {\bibinfo {volume} {595}},\ \bibinfo {pages} {317} (\bibinfo {year}
	{2008})}\BibitemShut {NoStop}%
\end{thebibliography}

%TC:endignore 

\end{document}